\documentclass[12pt, a4paper]{article}
\usepackage{amsmath,amssymb,amsfonts}
\usepackage{graphicx}

\begin{document}

\title{Casimir force induced on a plane by\\ an impenetrable flux tube of finite radius}

\author{Volodymyr M. Gorkavenko
\thanks{E-mail: gorka@univ.kiev.ua}\\
\it \small Department of Physics, Taras Shevchenko National
University of Kyiv,\\ \it \small 64 Volodymyrs'ka str., Kyiv
01601, Ukraine\\\phantom{11111111111}\\
Yurii A. Sitenko\thanks{E-mail: yusitenko@bitp.kiev.ua}, Olexander
B. Stepanov\thanks{E-mail:
\_\,pnd\_\,@ukr.net}\\
\it \small Bogolyubov Institute for Theoretical Physics,
\it \small National Academy of Sciences of Ukraine,\\
\it \small 14-b Metrologichna str., Kyiv 03680, Ukraine\\
\phantom{11111111111}}
\date{}

\maketitle

\begin{abstract}
A perfectly reflecting (Dirichlet) boundary condition at the edge of
an impenetrable magnetic-flux-carrying tube of nonzero transverse
size is imposed on the charged massive scalar matter field which is
quantized outside the tube on a plane which is transverse to the
tube. We show that the vacuum polarization effects outside the tube
give rise to a macroscopic force acting at the increase of the tube
radius (if the magnetic flux is held steady).

\end{abstract}


\section{Introduction}

Polarization of the vacuum of quantized matter fields under the
influence of boundary conditions was studied intensively over more
than six decades since Casimir \cite{Cas} predicted a force between
grounded metal plates: the prediction was that the induced vacuum
energy in bounded spaces gave rise to a macroscopic force between
bounding surfaces, see reviews in Refs.~\cite{Mil} and \cite{Bor}.
The Casimir force between grounded metal plates has now been
measured quite accurately and agrees with the theoretical
predictions, see, e.g. Refs.~\cite{Lam} and \cite{Bre}, as well as
other publications cited in Refs.~\cite{Mil} and \cite{Bor}.

In the present paper we consider the vacuum energy which is induced
by boundary conditions in space that is not bounded but, instead, is
not simply connected, being an exterior to a straight infinitely
long tube. This setup is inspired by the famous Aharonov-Bohm effect
\cite{Aha}, and we are interested in polarization of the vacuum
which is due to imposing a boundary condition at the edge of the
tube carrying magnetic flux lines inside itself; this may be denoted
as the Casimir-–Aharonov-Bohm effect (see also \cite{Sit}).

The vacuum polarization effects which are due to imposing boundary
conditions of various types at the cylindrical surfaces were
extensively discussed in the literature, see \cite{Raad} --
\cite{Cavero}. In general, the Casimir effect in the presence of a
single smooth object (cylinder or sphere) is rather different from
that in the presence of two disjoint ones (e.g., plates): new
divergences appear, and to tame them one has to sum contributions of
quantized matter from both sides of the boundary surface, still this
does not help in some cases to get rid completely of divergences,
see \cite{Bor} and references therein. In view of this, the
conventional prescription which is to subtract vacuum energy of
empty Minkowski space-time becomes insufficient for obtaining the
meaningful results. The author of ref.\cite{Schad} proposes to
define the Casimir energy for physical systems divided into classes:
the difference in vacuum energy of any two systems within the same
class should be finite, then the finite Casimir energy has the
universal interpretation as a vacuum energy with respect to the
vacuum energy of a certain reference system which is common for the
whole class. Following this proposition, we define a class of
physical systems corresponding to the charged massive scalar field
which is quantized outside an impenetrable tube with the magnetic
flux of different values; the case of zero flux can be chosen as the
reference system. As we shall show, the Casimir energy for this
class is unambiguous and finite.

The temporal component of the energy-momentum tensor for the
quantized charged scalar field $\Psi(x)$ in flat space-time is given
by expression
\begin{equation}\label{t00}
T_{00}(x)=\frac12\left[\partial_0\Psi^\dag,\partial_0\Psi\right]_+-\frac14\left[\partial_0^2\Psi^\dag,\Psi\right]_+
-\frac14\left[\Psi^\dag,\partial_0^2\Psi\right]_+-\left(\xi-\frac14\right){\mbox{\boldmath
$\nabla$}}^2\left[\Psi^\dag,\Psi\right]_+,
\end{equation}
where ${\mbox{\boldmath $\nabla$}}$ is the covariant spatial
derivative involving both affine and bundle connections and the
field operator in the case of a static background takes form
\begin{equation}\label{a11}
 \Psi(x^0,{\textbf{x}})=\sum\hspace{-1.4em}\int\limits_{\lambda}\frac1{\sqrt{2E_{\lambda}}}\left[e^{-{\rm i}E_{\lambda}x^0}\psi_{\lambda}({\bf x})\,a_{\lambda}+
  e^{{\rm i}E_{\lambda}x^0}\psi_{-\lambda}({\bf
  x})\,b^\dag_{\lambda}\right];
\end{equation}
 $a^\dag_\lambda$ and $a_\lambda$ ($b^\dag_\lambda$ and
$b_\lambda$) are the scalar particle (antiparticle) creation and
destruction operators satisfying commutation relations; wave
functions $\psi_\lambda(\textbf{x})$ form a complete set of
solutions to the stationary Klein-Gordon equation
\begin{equation}\label{a12}
 \left(-{\mbox{\boldmath $\nabla$}}^2  + m^2\right)  \psi_\lambda({\bf x})=E^2_\lambda\psi({\bf x}),
\end{equation}
 $m$ is the mass of the scalar
particle; $\lambda$ is the set of parameters (quantum numbers)
specifying the state; $E_\lambda=E_{-\lambda}>0$ is the energy of
the state; symbol
  $\sum\hspace{-1em}\int\limits_\lambda$ denotes summation over discrete and
  integration (with a certain measure) over continuous values of
  $\lambda$.

As is known for a long time \cite{Pen,Cher,Cal}, the energy-momentum
tensor depends on parameter $\xi$ which couples $\Psi$ to the scalar
curvature of space-time even in the case of the vanishing curvature,
see \eqref{t00}; conformal invariance is achieved in the limit of
vanishing mass $(m=0)$ at $\xi=(d-1)(4d)^{-1}$, where $d$ is the
spatial dimension. Consequently, the density of the induced vacuum
energy which is given formally by expression
\begin{equation}\label{a14}
   \varepsilon=\langle {\rm vac}|T_{00}(x)|{\rm vac} \rangle=\sum\hspace{-1.4em}\int\limits_{\lambda}E_\lambda\psi^*_\lambda(\textbf{x})\,\psi_\lambda(\textbf{x})-(\xi-1/4)
   {\mbox{\boldmath $\nabla$}}^2
  \sum\hspace{-1.4em}\int\limits_{\lambda}E^{-1}_\lambda\psi^*_\lambda(\textbf{x})\,\psi_\lambda(\textbf{x})
\end{equation}
depends on $\xi$ as well. This poses a question: whether physically
measurable effects (e.g. the Casimir force) can be dependent on
$\xi$?

In the present paper we are considering a static background in the
form of the cylindrically symmetric magnetic flux tube of finite
transverse size, hence the covariant derivative is $\mbox{\boldmath
$\nabla$}=\mbox{\boldmath $\partial$}-{\rm i}e {\bf V}$ with the
vector potential possessing only one nonvanishing component given by
\begin{equation}\label{3}
V_\varphi=\Phi/2\pi
\end{equation}
outside the tube; here $\Phi$ is the value of the magnetic flux and
$\varphi$ is the angle in  polar $(r,\varphi)$ coordinates on a
plane which is transverse to the tube. The Dirichlet boundary
condition at the edge of the tube $(r=r_0)$  is imposed on the
scalar field:
\begin{equation}\label{4}
\left.\psi_\lambda\right|_{r=r_0}=0,
\end{equation}
i.e. quantum matter is assumed to be perfectly reflected from the
thence impenetrable flux tube.

As we shall see, the vacuum energy that is induced outside the flux
tube on a plane which is transverse to the tube gives rise to a
macroscopic force acting at the increase of the tube radius if  the
magnetic flux is held steady. Although the induced vacuum energy
density is $\xi$-dependent, the Casimir force will be shown to be
independent of $\xi$.

\section{Vacuum energy density}

The solution to \eqref{a12} outside the magnetic flux tube can be
obtained in terms of the cylindrical functions. The formal
expression \eqref{a14} for the vacuum energy density has to be
renormalized by subtracting the contribution corresponding to the
zero flux. Restricting ourselves to a plane which is orthogonal to
the tube, we obtain (for details see \cite{newstring2}):
\begin{multline}\label{c2}
\varepsilon_{ren}=\frac1{2\pi}\left\{ \int\limits_0^\infty
  dk\,k\left(k^2+m^2\right)^{1/2}\left[S(kr,kr_0)-S(kr,kr_0)|_{\Phi=0}\right]-\right.\\\left.-(\xi-1/4)\triangle\int\limits_0^\infty
  dk\,k\left(k^2+m^2\right)^{-1/2}\left[S(kr,kr_0)-S(kr,kr_0)|_{\Phi=0}\right]\right\},
\end{multline}
 where
\begin{equation}\label{a29a}
 S(kr,kr_0)=\sum_{n\in\mathbb
 Z}\frac{\left[Y_{|n-e\Phi/2\pi|}(kr_0)J_{|n-e\Phi/2\pi|}(kr)-J_{|n-e\Phi/2\pi|}(kr_0)Y_{|n-e\Phi/2\pi|}(kr)\right]^2}{Y^2_{|n-e\Phi/2\pi|}(kr_0)+J^2_{|n-e\Phi/2\pi|}(kr_0)},
\end{equation}
$\mathbb{Z}$ is the set of integer numbers, $J_\mu(u)$ and
$Y_\mu(u)$ are the Bessel functions of order $\mu$ of the first and
second kinds, and $\triangle=\partial^2_r+r^{-1}\partial_r\,$ is the
 radial part of the Laplacian operator on the plane.

Owing to the infinite range of summation, the last expression is
periodic in flux $\Phi$ with a period equal to $2\pi e^{-1}$, i.e.
the London flux quantum (we use units $c=\hbar=1$). Our further
analysis concerns the case of $\Phi=(2n+1)\pi e^{-1}$ when each of
the integrals in \eqref{c2} is the most distinct from zero.
Introducing function \cite{newstring}
\begin{equation}\label{21}
G(kr,kr_0)=S(kr,kr_0)|_{\Phi=\pi e^{-1}}-S(kr,kr_0)|_{\Phi=0},
\end{equation}
we rewrite \eqref{c2} in the dimensionless form
\begin{equation}\label{1}
r^3\varepsilon_{ren}=\alpha_+(mr_0,mr)-(\xi-1/4)r^3\triangle
\frac{\alpha_-(mr_0,mr)}{r},
\end{equation}
where
\begin{equation}\label{c3ab}
\alpha_\pm(mr_0,mr)=\frac{1}{2\pi}\int\limits_0^\infty
dz\,z\left[z^2+\left(\frac{mr_0}\lambda\right)^2\right]^{\pm1/2}
G(z,\lambda z),
\end{equation}
and  $\lambda=r_0/r$ ($\lambda\in[0,1]$).

We follow the technique of numerical calculations developed in
\cite{newstring2,newstring} with some modifications, notably we
perform direct integration over consecutive periods of the
$G(z,\lambda z)$ function  using the Euler-Maclaurin integration
formula \cite{Euler}. This results in a sufficient decrease of the
computation time.

Thereafter we calculate $\alpha_+$ and $\alpha_-$ functions for the
case of $mr_0=10^{-3}$  at a set of different distances from the
axis of the tube. This allows us to obtain coefficients of the
interpolation function which is approximated in the form
\begin{equation}\label{m1}
\alpha_\pm(x_0,x)=\left[\pm e^{-2x}x^{1\mp1/2}\right]
\left[\left(\frac{x-x_0}{x}\right)^2\frac{P^\pm_3(x-x_0)}{x^3}\right]
\frac{Q^\pm_3(x^2)}{R^\pm_3(x^2)},\quad x>x_0,
\end{equation}
where $x=mr$, $x_0=mr_0$ and $P^\pm_n(y)$, $Q^\pm_n(y)$,
$R^\pm_n(y)$ --- are polynomials in $y$ of the $n$-th order with the
$x_0$-dependent coefficients. First factor in square bracket in
\eqref{m1} describes the large distance behavior  in the case of the
zero-radius tube (singular thread), second factor in  square bracket
is an asymptotics at small distances from the edge of the tube, and
the last factor in square bracket is the intermediate part. Since
the flux tube is impenetrable, the $\alpha_\pm$ functions vanish at
$x\leq x_0$. Behavior of the dimensionless $\alpha_\pm$ functions is
presented on Fig.1.

\begin{figure}[t]
\begin{center}
\includegraphics[width=85mm]{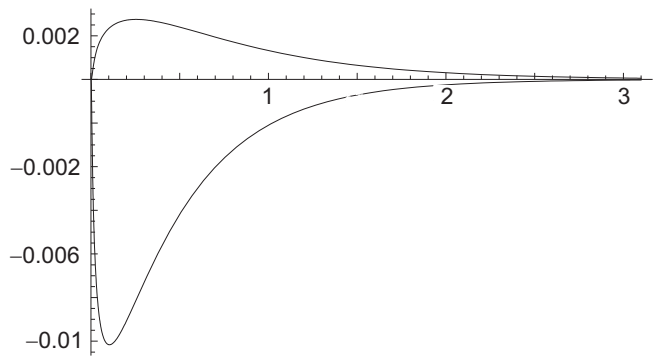}
\end{center}
\caption{Behavior of the $\alpha_+(x_0,x)$ (positive) and the
$\alpha_-(x_0,x)$ (negative) functions for the case of
$x_0=10^{-3}$. The variable $x$ $(x>x_0)$ is along the abscissa
axis.\label{fig1}}
\end{figure}

For the $\alpha_+$ function  we estimate the relative error of the
obtained result as $0.1\%$. It should be noted that nearly 95 \% of
the integral value  is obtained by direct calculation and only
nearly 5\% is the contribution from the interpolation. The
integration in the case of the $\alpha_-$ function is performed more
quickly and with a higher accuracy, as compared to the case of the
$\alpha_+$ function, because the former tends to zero more rapidly
at large distances. In this case the contribution from the
interpolation can be estimated as $10^{-3}\%$ from the total value.

\begin{figure}[b]
\begin{center}
\includegraphics[width=155mm]{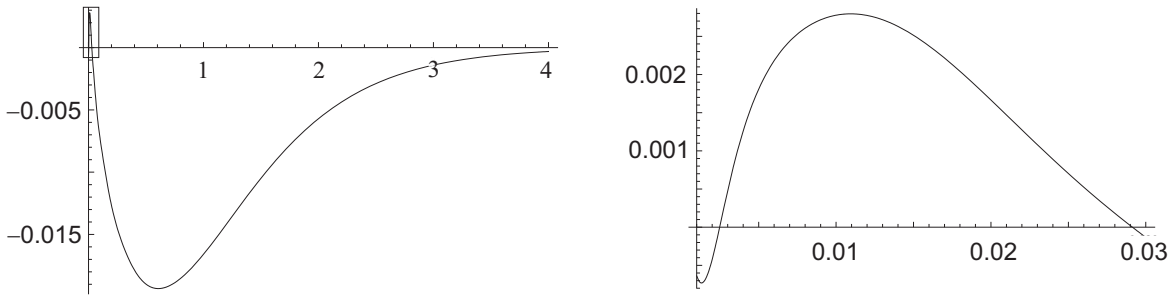}
\end{center}
\caption{Behavior of the $\tilde\alpha_-(x_0,x)$  function for the
case of  $x_0=10^{-3}$. The region in a rectangle on the left figure
is seen in the scaled-up form on the right figure.
 The variable $x$ $(x>x_0)$ is along the abscissa axis.\label{fig2}}
\end{figure}

We define function \cite{newstring2}
\begin{equation}\label{m2}
\tilde
\alpha_-(x_0,x)=r^3\triangle\frac{\alpha_-(x_0,x)}{r}=\alpha_-(x_0,x)-x\frac{\partial
\alpha_-(x_0,x)}{\partial x}+x^2\frac{\partial^2
\alpha_-(x_0,x)}{\partial x^2}
\end{equation}
and present its behavior on Fig.2.

We construct the dimensionless vacuum energy density  at different
values of the coupling to the space-time curvature scalar ($\xi$):
\begin{equation}\label{m3}
r^3\varepsilon_{ren}=\alpha_+(x_0,x)-(\xi-1/4)\tilde\alpha_-(x_0,x)
\end{equation}
and present its behavior in the case of $mr_0=10^{-3}$ on Fig.3.

The  behavior of the induced vacuum energy density as the radius of
the tube tends to zero is of primary interest. To do a numerical
calculation  at $x_0<10^{-3}$  needs a long computational time and
is a rather complicated task. Nevertheless, we can make some general
conclusions regarding the case of small $x_0$.

It seems plausible that this case with the decrease of the tube
radius becomes more similar to the case of the tube of zero radius
(singular thread) see, e.g., Fig.4. However there are some
peculiarities in the behavior in the vicinity of the tube. To
discuss them, let us first recall the exact expressions
corresponding to the case of the singular magnetic thread (see
\cite{our2}):

\begin{figure}[t]
\begin{center}
\includegraphics[width=160mm]{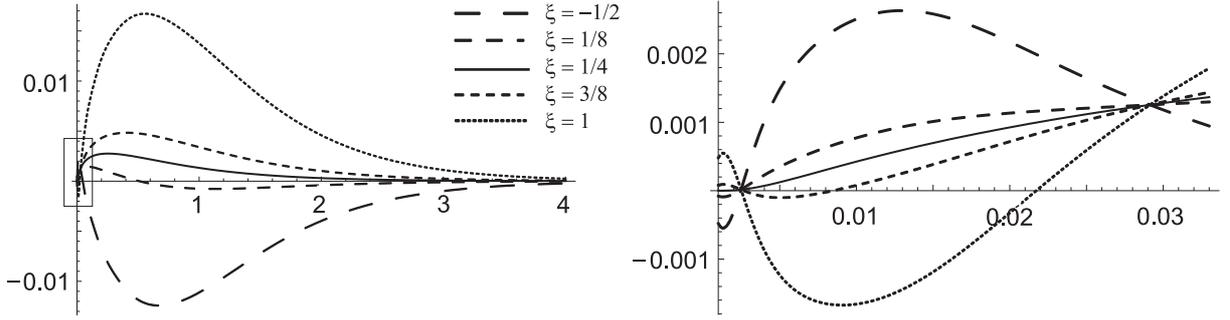}
\end{center}
\caption{The dimensionless vacuum energy density
$r^3\varepsilon_{ren}(x_0,x)$ at different values of the coupling to
the space-time curvature scalar for the case of $x_0=10^{-3}$. The
region in a rectangle on the left figure is seen in the scaled-up
form on the right figure.
 The variable $x$ $(x>x_0)$ is along the abscissa axis.\label{fig3}}
\end{figure}

\begin{figure}[t]
\begin{center}
\includegraphics[width=150mm]{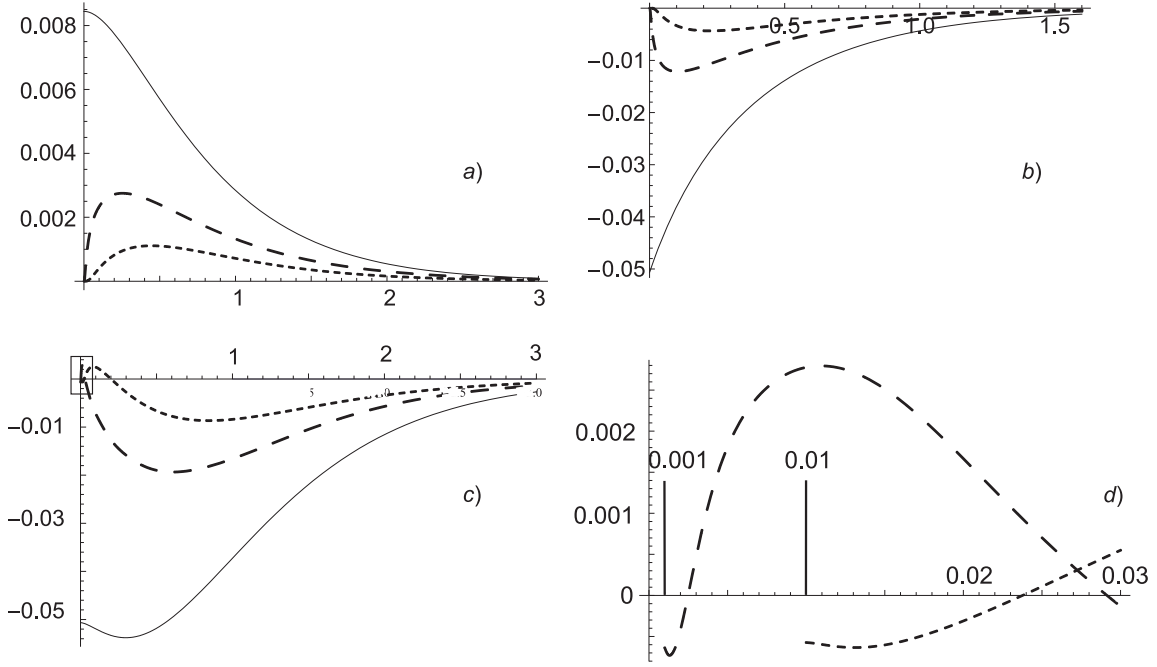}
\end{center}
\caption{The constituents of the dimensionless vacuum energy
density: \textit{a}) $\alpha_+$, \textit{b}) $\alpha_-$, \textit{c})
$\tilde \alpha_-$ for the case of $x_0=10^{-2}$ (dotted line),
$10^{-3}$ (dashed line). The region in a rectangle on the
\textit{c}-figure is seen in the scaled-up form on the
\textit{d}-figure. The behavior of the corresponding functions for
the case of a singular magnetic thread is presented by solid line.
The variable $x$ $(x>0)$ is along the abscissa axis.\label{fig4}}
\end{figure}

\begin{align}\label{singa}
&\alpha_+(0,x)=\frac{x^3}{3\pi^2}\left\{\frac\pi2-2x
K_0(2x)-K_1(2x)+\frac{K_2(2x)}{2x}-\right.\nonumber\\
&\hspace{16.9em}\left.\vphantom{\frac12}-\pi x
\left[K_0(2x)L_{1}(2x)+K_1(2x)L_{0}(2x)\right]\right\},\\
&\alpha_-(0,x)=\frac{x}{\pi^2}\left\{\frac\pi2-2x
K_0(2x)-K_1(2x)-\pi
x \left[K_0(2x)L_{1}(2x)+K_1(2x)L_{0}(2x)\right]\right\},\\
&\tilde\alpha_-(0,x)=-\frac{x}{\pi^2}[2x K_0(2x)+K_1(2x)],
\end{align}
where $K_\nu(u)$ and $L_\nu(u)$ are the Macdonald and the modified
Struve functions of  order $\nu$. Consequently, in the vicinity of a
thread one gets
\begin{align}\label{singasmall1}
&\alpha_+(0,x)=\frac{1-3x^2}{12\pi^2},\quad x\ll 1\\
&\alpha_-(0,x)=-\frac{
1 - \pi x + (3  - 2 \gamma - 2 \ln x)x^2}{2\pi^2}+O(x^3),\quad x\ll 1,\label{singasmall2}\\
&\tilde\alpha_-(0,x)=-\frac1{2\pi^2}+\frac{1+2\gamma+2\ln
x}{2\pi^2}\,x^2+O(x^3),\quad x\ll 1,\label{singasmall3}
\end{align}
where $\gamma$ is the Euler constant. Using the latter relations, we
get asymptotics of the renormalized vacuum energy density at small
distances from the singular magnetic thread
\begin{equation}\label{g8}
r^3\varepsilon_{ren}^{sing}=\frac1{12\pi^2}-\frac{x^2}{4\pi^2}-\left(\xi-\frac14\right)\left(-\frac{1}{2\pi^2}+\frac{1
+ 2 \gamma + 2 \ln x }{2\pi^2}x^2\right)+O(x^3),\quad x\ll1.
\end{equation}

In contrast to \eqref{singasmall1} and \eqref{singasmall2}, the
$\alpha_\pm(x_0,x)$ functions in the case of nonzero radius are
vanishing quadratically in the vicinity of the tube, see
\cite{newstring2},
\begin{equation}\label{g8a}
\left.\alpha_\pm(x_0,x)\right|_{x\rightarrow x_0}\sim
O\left[(x-x_0)^2\right].
\end{equation}
To be more precise, we assume the asymptotics in the form, cf.
\eqref{m1},
\begin{equation}\label{alpm}
\alpha_\pm(x_0,x)=\pm\frac{(x-x_0)^2}{x^2}f_\pm(x_0,x),
\end{equation}
then one gets
\begin{multline}\label{dop7}
\tilde\alpha_-(x_0,x)=-(x-x_0)^2\frac{\partial^2}{\partial
x^2}f_-(x_0,x)+\left(1-6\frac{x_0}{x}+5\frac{x_0^2}{x^2}\right)x\frac{\partial}{\partial
x}f_-(x_0,x)-\\-\left(1-8\frac{x_0}{x}+9\frac{x_0^2}{x^2}\right)f_(x_0,x),
\end{multline}
with $\tilde\alpha_-(x_0,x_0)=-2f_-(x_0,x_0)$.

\begin{figure}[t]
\begin{center}
\includegraphics[width=150mm]{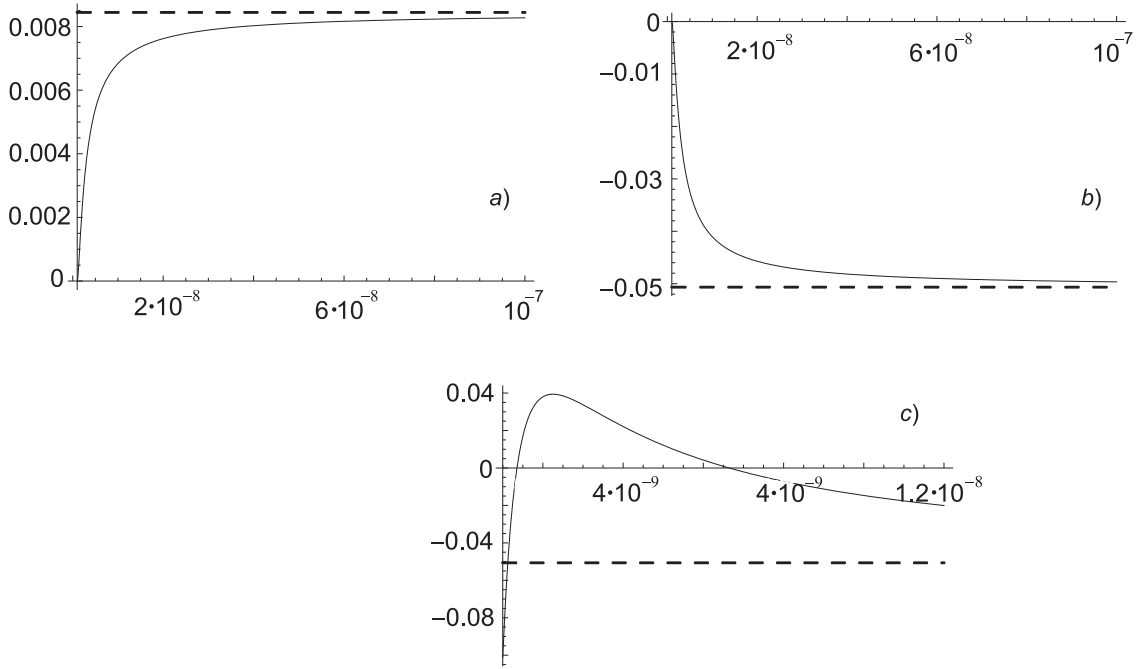}
\end{center}
\caption{The expected behavior of the constituents of the
dimensionless vacuum energy density at small distances from the
tube: \textit{a}) $\alpha_+$, \textit{b}) $\alpha_-$, \textit{c})
$\tilde \alpha_-$  for the case of $x_0=10^{-9}$ (solid line). The
behavior of the corresponding functions for the case of a singular
magnetic thread is presented by dashed line. The variable $x$
$(x>x_0)$ is along the abscissa axis.\label{fig5}}
\end{figure}

The $f_\pm(x_0,x)$ functions are adjusted as
\begin{align}\label{001}
& f_+(0,x)=\frac{
1-3x^2}{12\pi^2},\quad x\ll1,\\
& f_-(0,x)=\frac{ 1 - \pi x + (3  - 2 \gamma - 2 \ln
x)x^2}{2\pi^2},\quad x\ll1\label{002};
\end{align}
consequently, one gets
\begin{multline}\label{003}
\left.\vphantom{\frac12}\tilde\alpha_-(x_0,x)\right|_{\!\!\!\scriptsize\begin{array}{l}
x_0\rightarrow 0\\ x\rightarrow x_0\end{array}}
=-\frac1{2\pi^2}+\frac{1+2\gamma+2\ln x}{2\pi^2}\,x^2+\frac{4-\pi
x}{\pi^2x}\,x_0+\\+\frac{-9 + 4 \pi x - 7 x^2 + 2 \gamma x^2 + 2 x^2
\ln x}{2\pi^2 x^2}\,x_0^2.
\end{multline}

The asymptotical behavior of the $\alpha_\pm$ and $\tilde\alpha_-$
functions with the use of \eqref{alpm} -- \eqref{003}  is presented
on Fig.5 for the case of a sufficiently small value of $x_0$. As one
can see, this behavior is quite similar to the behavior for the case
of $x_0=10^{-3}$, compare with Fig.2 and Fig.4. It should be noted
that the $f_\pm(x_0,x)$ functions depend strongly on $x_0$.

\section{Total vacuum energy and the Casimir force}

The total vacuum energy which is induced on a plane outside the
magnetic flux tube of finite radius is
\begin{equation}\label{g3a}
E\equiv\int\limits_{0}^{2\pi}d\varphi\int\limits_{r_0}^\infty
\varepsilon_{ren} \,r dr=2\pi m \left[\int\limits_{x_0}^\infty
\frac{\alpha_+(x_0,x)}{x^2}\,dx-\left(\xi-\frac14\right)\int\limits_{x_0}^\infty
\frac{\tilde\alpha_-(x_0,x)}{x^2}\,dx\right].
\end{equation}
In view of the relation
\begin{equation}\label{m5}
\int\limits_{x_0}^\infty \frac{\tilde\alpha_-(x_0,x)}{x^2}\,
dx=\left.\left[\frac{\alpha_-(x_0,x)}{x}-\frac{\partial\alpha_-(x_0,x)}{\partial
x}\right]\right|_{x=x_0}
\end{equation}
which follows from \eqref{m2}, and relations \eqref{alpm} and
\eqref{002}, we conclude that the vacuum energy is independent of
the coupling to the space-time curvature scalar ($\xi$):
\begin{equation}\label{dop1}
E=2\pi m \int\limits_{x_0}^\infty \frac{\alpha_+(x_0,x)}{x^2}\,dx.
\end{equation}
This is in contrast to the case of the singular magnetic thread,
when the total induced vacuum energy is divergent and
$\xi$-dependent (see \cite{our2}):
\begin{equation}\label{dop2}
E^{sing}\equiv\int\limits_{0}^{2\pi}d\varphi\int\limits_{0}^\infty
\varepsilon_{ren}^{sing} \,r dr\sim
4m\left(\xi-\frac1{12}\right)\int\limits_0\frac{dx}{x^2}.
\end{equation}
It is curious that the vacuum energy in this case is finite at
$\xi=1/12$, being equal to
\begin{multline}\label{dop3}
\left.E^{sing}\right|_{\xi=1/12}=\frac{2m}{3\pi}\int\limits_0^\infty\left\{\frac\pi2-\left(2x+\frac1{2x}\right)K_0(2x)-K_1(2x)-\right.\\
-\left.\vphantom{\frac12}\pi
x\left[K_0(2x)L_1(2x)+K_1(2x)L_0(2x)\right]\right\}x\,dx=-0.01989\times
2\pi m.
\end{multline}

Although vacuum energy $E$ \eqref{dop1} is finite, its absolute
value grows infinitely as $x_0$ tends to zero (see \eqref{alpm} and
\eqref{001}):
\begin{equation}\label{dop3}
\left.E\right|_{x_0\rightarrow0}=m\left[\frac{1}{18\pi
x_0}-\frac{x_0}{\pi}\ln x_0+O(x_0^3)\right],
\end{equation}
which is in accordance with the divergence of the vacuum energy in
the case of the singular magnetic thread. To be more precise,
relation \eqref{m5} fails to yield zero in the case $x_0=0$, and,
therefore, the divergence of the vacuum energy in the latter case
becomes $\xi$-dependent.

\begin{figure}[t]
\begin{center}
\includegraphics[width=90mm]{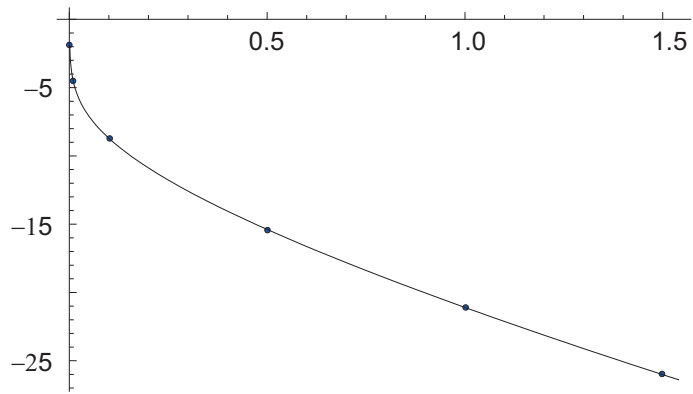}
\end{center}
\caption{The total vacuum energy   as a function of the tube radius
in the range $10^{-3}<x_0<3/2$. The variable $x_0$ is along the
abscissa axis, the value of $\ln \frac{E}{2\pi m}$ is along the
ordinate axis. Solid line interpolates the dots that have been
calculated. \label{fig6}}
\end{figure}

We present the values of vacuum energy $E$ \eqref{dop1} for several
values of the tube radius in the Table:
\begin{center}
\begin{tabular}{|c|c|c|c|c|c|c|}
\hline $x_0$ & 3/2 & 1 & $1/2$ & $10^{-1}$ & $10^{-2}$ & $10^{-3}$\\
\hline $E/(2\pi m)$ & $5.013\cdot 10^{-12}$ & $6.944\cdot10^{-10}$ &
$2.068\cdot10^{-7}$ & $1.65\cdot 10^{-4}$ &
$0.0106$ & $0.1486$\\
\hline
\end{tabular}
\end{center}
 These results are also given
on Fig.6 in logarithmic scale, where the dots corresponding to the
data in the Table are joined with the help of the interpolation
function,
\begin{equation}\label{dop4}
\eta(x_0)=\ln \frac{E}{2\pi m},
\end{equation}
which can be taken in the form
\begin{equation}\label{g2}
\eta(x_0)=a_0+\sum_i a_i x_0^{b_i}-\left(1+\sum_j c_j
x^{d_j}_0\right)\ln x_0.
\end{equation}
where $b_i$ and $d_j$ are the positive adjustable constants.

To change the radius of the  magnetic flux tube one has to apply a
work that is equal to the change of the total vacuum energy which is
induced outside the tube. In the case of the infinitely  small
change of the radius one has
\begin{equation}\label{g4}
\Delta E=2\pi p\, r_0 \Delta r_0 ,
\end{equation}
where $p$ can be interpreted as the vacuum pressure which acts from
the outside to the inside of the tube
\begin{equation}\label{g5}
p(x_0)=\frac{1}{2\pi r_0}\frac{d E}{d r_0}=m^3
\frac{e^{\eta(x_0)}}{x_0}\frac{d \eta(x_0)}{d x_0}.
\end{equation}
This results in the Casimir force acting from the inside to the
outside of the tube
\begin{equation}\label{g6}
F(x_0)=-2\pi r_0  p(x_0)=- 2\pi m^2 e^{\eta(x_0)}\frac{d
\eta(x_0)}{d x_0}.
\end{equation}
The behavior of the Casimir force is presented on Fig.7.

As one can see, the Casimir force tends to increase the radius of
the tube and to minimize the induced vacuum energy of the quantized
scalar field. Certainly, our conclusion is obtained under the
assumption that the magnetic flux inside the impenetrable tube
remains invariable with the variation of the tube radius.

\section{Summary}

\begin{figure}[t]
\begin{center}
\includegraphics[width=80mm]{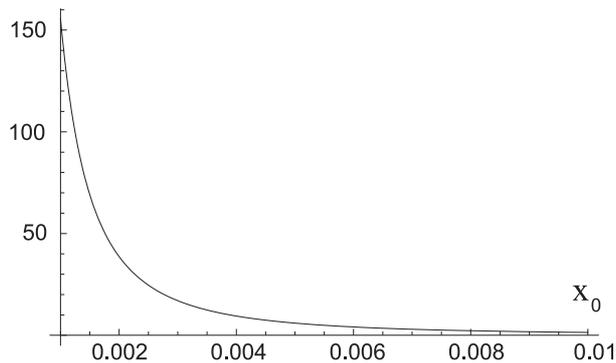}
\end{center}
\caption{The Casimir force as a function of the tube radius. The
variable $x_0$ is along the abscissa axis, the value of the
dimensionless Casimir force $\frac{F(x_0)}{2\pi m^2}$ is along the
ordinate axis.\label{fig7}}
\end{figure}

In the present paper we consider the vacuum polarization effects
which are induced in scalar matter by imposing a perfectly
reflecting (Dirichlet) boundary condition at the edge of an
impenetrable finite-radius tube which carries magnetic flux lines
inside itself. Restricting ourselves to a plane which is orthogonal
to the tube, we define the induced vacuum energy density, see
\eqref{c2} and \eqref{a29a}, and analyze numerically its behavior as
a function of the distance from the tube for the magnetic flux equal
to half of the London flux quantum ($\Phi=\pi e^{-1}$), the tube
radius equal to $r_0=10^{-3}m^{-1}$ and different values of the
coupling to the space-time curvature scalar ($\xi$), see Fig.3. The
emergence of the energy density, as well as of other components of
the energy-momentum tensor, in the vacuum can lead to various
semiclassical gravitational effects which were estimated under the
neglect of the tube radius in \cite{GorViz}.

The present paper summarizes and extends our previous study in
\cite{newstring2,newstring}, and this allows us to draw conclusions
about the behavior of the total induced vacuum energy, i.e. the
density integrated over the whole plane, and the Casimir force as
functions of the tube radius. We find that the total induced vacuum
energy is finite and independent of $\xi$, as long as the tube
radius is taken into account. Although the values of the total
induced vacuum energy are negligible for $r_0\sim m^{-1}$ (see also
\cite{newstring}) being of order $10^{-10}\times 2\pi m$, they are
of order $10^{-1}\times 2\pi m$ for $r_0\sim 10^{-3}m^{-1}$, see the
Table and Fig.6. The induced vacuum energy gives rise to the Casimir
force which is directed from the inside to the outside of the tube.
The force acts at the increase of the tube radius and the decrease
of the induced vacuum energy, if the magnetic flux is held
constant\footnote{As to the energy stored inside the tube, it is the
purely classical energy of the magnetic field. Its behavior at the
increase of the tube radius as the magnetic flux is held constant
can be different depending on the details of the magnetic field
configuration. Mild assumptions as to the smoothness of the
configuration yield that the energy is either constant or decreasing
at most as $\sim r_0^{-2}$.}. The force takes considerable values at
small values of the tube radius and actually disappears otherwise:
it is, e.g., $10^2\times 2\pi m^2$ at $r_0\sim 10^{-3}m^{-1}$ and
$10^{-3}\times 2\pi m^2$ at $r_0\sim 10^{-1}m^{-1}$. The behavior of
the force as a function of the tube radius is illustrated by Fig.7.

It should be noted that in our case the Casimir force is caused by
boundary conditions imposed at the boundary enclosing a magnetic
flux. The force is periodic in the flux value with a period equal to
the London flux quantum, attaining its maximal value at
$\Phi=(2n+1)\pi e^{-1}$ and vanishing at $\Phi= 2n\pi e^{-1}$ ($n\in
\mathbb Z$).

\section*{Acknowledgments}

The work was partially supported by special program "Microscopic and
phenomenological models of fundamental physical processes in micro-
and macroworld" of the Department of Physics and Astronomy of the
National Academy of Sciences of Ukraine.

\end{document}